

\font\titolino=cmbx10
\font\tsnorm=cmr10
\font\tscors=cmti10

\font\tscorsp=cmti9
\magnification=1200

\hsize=148truemm
\hoffset=10truemm
\parskip 3truemm plus 1truemm minus 1truemm
\parindent 8truemm
\newcount\notenumber

\def\PRD{{\tscors Phys. Rev. D }}
\def\PRL{{\tscors Phys. Rev. Lett. }}
\def\NP{{\tscors Nucl. Phys. }}

\def\IJMPA{{\tscors Int. J. Mod. Phys. A  }}
\def\MPLA{{\tscors Mod. Phys. Lett. A  }}
\def\CQG{{\tscors Class. Quantum Grav. }}

\def\E{Ein\-stein}
\def\Sc{Sch\-warz\-sch\-ild}
\def\Schr{Sch\-r\"o\-din\-ger}
\def\K{Ku\-cha\v r}
\def\bh{bla\-ck \-ho\-le}
\def\bhs{bla\-ck \-ho\-les}

\def\BHs{Bla\-ck \-Ho\-les}
\def\d{\partial}

\def\wh{worm\-ho\-le}

\def\Whs{Worm\-ho\-les}
\def\ks{Kan\-tow\-ski\--Sa\-chs}

\def\note{\advance\notenumber by 1 \footnote{$^{\the\notenumber}$}}
\def\ref#1{\medskip\everypar={\hangindent 2\parindent}#1}
\def\beginref{\begingroup
\bigskip
\leftline{\titolino References.}
\nobreak\noindent}
\def\endref{\par\endgroup}
\def\beginsection #1. #2.
{\bigskip
\leftline{\titolino #1. #2.}
\nobreak\noindent}

\def\beginack
{\bigskip
\leftline{\titolino Acknowledgments}
\nobreak\noindent}

\nopagenumbers
\null
\vskip 5truemm
\rightline {DFTT 64/94}
\rightline{November 26, 1994}
\vskip 15truemm
\centerline{\titolino HAMILTONIAN FORMALISM FOR BLACK HOLES}
\bigskip
\centerline{\titolino AND QUANTIZATION}
\vskip 10truemm
\centerline{\tsnorm Marco Cavagli\`a$^{(a),(d)}$,
Vittorio de Alfaro$^{(b),(d)}$ and Alexandre T. Filippov$^{(c),(d)}$}
\bigskip
\centerline{$^{(a)}$\tscorsp SISSA - International School for Advanced
Studies,}
\smallskip
\centerline{\tscorsp Via Beirut 2-4, I-34013 Trieste, Italy.}
\bigskip
\centerline{$^{(b)}$\tscorsp Dipartimento di Fisica
Teorica dell'Universit\`a di Torino,}
\smallskip
\centerline{\tscorsp Via Giuria 1, I-10125 Torino, Italy.}
\bigskip
\centerline{$^{(c)}$\tscorsp Joint Institute for Nuclear Research}
\smallskip
\centerline{\tscorsp R-141980 Dubna, Moscow Region, RUSSIA.}
\bigskip
\centerline{$^{(d)}$\tscorsp INFN, Sezione di Torino, Italy.}
\vskip 15truemm
\centerline{\tsnorm ABSTRACT}
\begingroup\tsnorm\noindent
Starting from the Lagrangian formulation of the \E\ equations for the
vacuum static spherically symmetric metric, we develop a canonical
formalism in the radial variable $r$ that is time--like inside the \Sc\
horizon.  The \Sc\ mass turns out to be represented by a canonical
function that commutes with the $r$--Hamiltonian. We investigate the
Whee\-ler--DeWitt quantization and give the general representation for
the solution as superposition of eigenfunctions of the mass operator.
\vfill
\leftline{\tsnorm PACS: 04.20.Fy, 04.60.Ds, 04.70.-s.\hfill}
\smallskip
\hrule
\noindent
\leftline{E-Mail: CAVAGLIA@TSMI19.SISSA.IT\hfill}
\leftline{E-Mail: VDA@TO.INFN.IT\hfill}
\leftline{E-Mail: FILIPPOV@THSUN1.JINR.DUBNA.SU\hfill}
\endgroup
\vfill
\eject
\footline{\hfill\folio\hfill}
\pageno=1
%
\beginsection 1. Introduction.
Recently the dynamics of primordial \Sc\ \bhs\ has been cast in canonical
formalism and the quantization procedure has been discussed [1]. A
complete bibliography that covers the history of the subject is also
contained there. Indeed, the Hamiltonian formalism is a fundamental key
to obtain a quantum description of a gravitational system and a great
deal of work has been devoted to the construction of a canonical
formalism for the classical \bh\ solutions (see also [2,3]).

In the present paper we derive the canonical formalism for the vacuum
static spherically symmetric metric in a simple direct way by foliation
in the coordinate $r$. Classically the general vacuum spherically
symmetric solution of the \E\ equations is locally isometric to the \Sc\
metric. In order to obtain a Hamiltonian description of the \Sc\ metric
we start from the general static spherically symmetric line element [4]
$$ds^2=-a(r)dt^2+N(r)dr^2+2B(r)dtdr+b(r)^2d\Omega^2,\eqno(1.1)$$
where $a$, $B$, $N$ and $b$ are real functions of $r$ and
$d\Omega^2=d\theta^2+\sin^2\theta d\varphi^2$ is the metric of the
two--sphere.

Usually, redefining the coordinate time and fixing $b=r$, (1.1) is cast
in the form
$$ds^2=-A(r)dt^2+C(r)dr^2+r^2d\Omega^2,\eqno(1.2)$$
where $r$ is now the ``area coordinate'' since the area of the two-sphere
of radius $r$ is $4\pi r^2$. One is then left with two functions $A(r)$
and $C(r)$ that can be determined by the \E\ equations. The line element
(1.2) is the ``standard form'' of the general static isotropic metric
(1.1) [4].

The line element (1.1) will be our starting point for a canonical
treatment, formulated in the coordinate $r$. We are of course aware that
the line element (1.1) does not cover the complete spacetime since it
describes only a half of the Kruskal--Szekeres plane and pure
$r$--coordinate transformations do not lead to a complete covering of the
Kruskal--Szekeres manifold starting from the metric (1.1). In spite of
this, the analysis of reparametrizations from this point of view may lead
to interesting consequences.  Indeed, later we will consider a formal
$r$--quantization scheme and investigate the ensuing Wheeler -- DeWitt
(WDW) equation [5,6].

Since the metric tensor in (1.1,2) does not depend on $t$, no
$t$--differentiation appears in the expression of the minisuperspace
action derived from (1.1); starting from the Lagrangian we may develop a
formal Hamiltonian scheme in the variable $r$ and obtain the
corresponding $r$--super Hamiltonian $H(a,p_a,b,p_b)$ after having
introduced the $r$--conjugate momenta $p_a$ and $p_b$.

Note that there is a range where $r$ is a timelike variable. The signs of
$N$ and $a$ are the key. In fact, inside the \Sc\ horizon of the \bh\ the
area coordinate $r$ in (1.2) is a time variable while $t$ is spacelike,
and our formalism is a true canonical motion in time. In the range where
$r$ is timelike, $H$ generates the dynamics and plays the role of the
usual ADM Hamiltonian; in general the $r$--super Hamiltonian is related
to the reparametrizations of the variable $r$. In the metric (1.1)
$\sqrt{|N|}$ plays essentially the role of the ADM lapse function with
respect to the $r$--slicing [7]. Since we must allow for negative values
of $N(r)$, we need a slight modification of the ADM formalism, similar to
what has been done, for instance continuing from a Lorentzian to an
Euclidean signature (see e.g. [7]).

A single Lagrange multiplier imposes the constraint of va\-ni\-shing of
the $r$--su\-per Ha\-mil\-to\-nian
$$H(a,p_a,b,p_b)=0,\eqno(1.3)$$
so of course the Hamiltonian ``$r$--dynamics'' is generated by a
constraint that is quadratic in the momenta, as predicted by the ADM
canonical formalism. It is easy to check that this formalism is
equivalent to the \E\ equations for the static solution.

The canonical formalism allows for an interesting algebraic structure of
constants of the motion: in particular we will see that the \Sc\ mass is
expressed by a constant canonical quantity, of course gauge invariant.

The constraint equation $H=0$ is independent of $r$, indeed there has
been no gauge fixing and $r$ is not determined. The identification of $r$
should be obtained by connecting it to the canonical coordinates of the
problem (gauge fixing). This procedure can be carried on by the method
proposed in [8] for quantum cosmological models. We defer to further
study the analysis of gauge fixing and quantization in the reduced space.

We will investigate the quantization of the system by the method of
enforcing the condition $H=0$ as an operator condition over wave
functions (WDW equation). We find the form of the general solution of the
equation diagonalizing the \Sc\ mass operator and a commuting operator.
The solutions have an oscillatory behaviour in the classically allowed
regions and an exponential behaviour in the classically forbidden ones.

Thus in this approach the mass plays the role of the quantum number
determining the wave function; in this respect our result is in agreement
with the conclusions obtained in [1].

The outline of the paper is as follows. In the next section we discuss
the classical $r$--Lagrangian and $r$--Hamiltonian formalisms for the
metric (1.1).  In section 3 we integrate the infinitesimal gauge
transformations and obtain the entire group. We identify the gauge
invariant quantities and discuss their algebra. Section 4 is devoted to
the study of the WDW equation.
\beginsection 2. Lagrangian formulation.
Our starting point is the line element (1.1) where the Lagrangian
coordinates $a$, $b$, $B$, $N$ are functions of $r$. As mentioned in the
introduction, changes of sign in the metric coefficients $a$ and $N$ are
allowed (note that the signature is Minkowskian over the whole manifold:
for instance, if $B=0$, $aN>0$). $r$ can be a timelike coordinate and $t$
spacelike over part of the manifold, so it is a matter of preference to
define {\tscors a priori} $t$ or $r$ as the timelike variable. Hence, we
develop a formal canonical structure in $r$ in which the $r$--super
Hamiltonian $H$ is a generator of gauge canonical transformations that
correspond to reparametrizations of the $r$ coordinate in the Lagrangian
formulation (and thus in the region where $r$ is timelike it generates
the dynamics). Hence it seems worthwile to study in detail this
$r$--canonical structure.

Let us consider the line element (1.1) that corresponds essentially to
use a Gaussian normal system of coordinates with respect to the
three--surface $(t,\theta,\phi)$, i.e. to perform the 3+1 slicing with
respect to the $r$ coordinate. As remarked in the introduction, looking
at (1.1) one realizes that the variable $\sqrt{|N(r)|}$ plays the role of
the $r$--lapse function in our foliation [7]. The \E--Hilbert action
$$S={1\over 16\pi G}\int_{V_4} d^4x \sqrt{-g} R -{1\over 8\pi G}
\int_{\d V_4} d^3x~\sqrt{h}~{\bf K}\eqno(2.1)$$
can be cast in the form
$$S=\int_{t_1}^{t_2} dt\int_{r_1}^{r_2} dr L(a,b,\Delta),\eqno(2.2)$$
where
$$L=2\sqrt{\Delta}\left({a'bb'\over\Delta}+
{ab'^2\over\Delta}+1\right).\eqno(2.3)$$
(primes denote differentiation with respect to $r$). In (2.3) we have set
$4G=1$ and $\Delta$ is given by
$$\Delta(r)=aN+B^2.\eqno(2.4)$$
Eq. (2.3) requires that $\Delta>0$ and from (2.4) the signature of (1.1)
is Minkowskian for any value of $r$. From (2.3) the \E\ equations of
motion can be recovered considering formally $a(r)$, $N(r)$, $B(r)$ and
$b(r)$ as Lagrangian coordinates evolving in $r$.  Of course,
$\sqrt\Delta$ acts as a Lagrange multiplier (and we still have the
freedom of choosing $B(r)$ or $N(r)$). From the vacuum \E\ equations
derived from (2.1), or directly from (2.3), one obtains
$$\eqalignno{&\Delta=k^2b'^2,&\hbox{(2.5a)}\cr\cr
&a=k^2\biggl(1-{2M\over b}\biggr),&\hbox{(2.5b)}\cr}$$
where $k$ and $M$ are two integration constants. Since the metric is
$t$--independent, we can arbitrarily rescale $t$ in (1.1). This
corresponds essentially to fix $k$ in (2.5), so we can set $k=1$; then
the metric coincides with the standard \Sc\ form. $M$ is the \Sc\ mass.
Eqs. (2.5) will be useful for comparison with the Hamiltonian formalism
that will be developed below. Note that the Lagrange multiplier
$\sqrt{\Delta}$ can be arbitrarily fixed; furthermore, since $\Delta$ is
related to $N$ and $B$ by eq. (2.4), also $N$, or $B$, can be arbitrarly
chosen; these two choices correspond to the freedom in the definition of
$t$ and $r$ in the line element (1.1). For instance, the choice
$\Delta=1$ corresponds to the area gauge since from (2.5) we obtain
$r=b$:
$$ds^2=-\left(1-{2M\over r}\right)dt^2+N(r)dr^2\pm
2\left[1-\left(1-{2M\over r}\right)N(r)\right]^{1/2}
dt~dr+r^2d\Omega^2.\eqno(2.6)$$
The line element (2.6) corresponds to the standard form of the \Sc\
solution for $N(r)=(1-2M/r)^{-1}$, to the Eddington--Finkelstein metric
for $N=1+2M/r$ and to the line element of ref. [2] choosing $N=1$.

Let us now set up the Hamiltonian formalism in $r$. We introduce the
$r$--conjugate momenta as
$$\eqalignno{&p_a={2bb'\over\sqrt\Delta},&\hbox{(2.7a)}\cr\cr
&p_b={2\over\sqrt\Delta}(a'b+2ab'),&\hbox{(2.7b)}\cr}$$
and by the usual Legendre transformation we obtain the density of the
action (with respect to the coordinate $t$)
$${\cal S}=\int_{r_1}^{r_2} dr\biggl\{ {1 \over 2}
(a'p_a + b'p_b - ap'_a - bp'_b)-lH\biggr\}.\eqno(2.8)$$
$H$ is the \Sc\ $r$--super Hamiltonian
$$H=p_a(bp_b-ap_a)-4b^2,\eqno(2.9)$$
and
$$l={\sqrt\Delta\over 2b^2}\eqno(2.10)$$
has been chosen as Lagrange multiplier. Note that the Legendre
transformation used to write (2.8) is singular for $b=0$, but not for
$a=0$. As a consequence of (2.8) we have the constraint
$$H=0.\eqno(2.11)$$
This constraint expresses the invariance under $r$--reparametrization and
inside the region where $r$ is timelike it generates the dynamics.

Eqs. (2.1) -- (2.11) can be easily extended to the Reissner --
Nordstr\"om (RN) case, i.e. to a static electrically charged \bh. Let us
consider a radial electric field whose potential 1-form is
$$A = A(r) dt \eqno(2.12)$$
(this Ansatz was used in [9,10] for the discussion of Euclidean
electromagnetic \bh s). Adding to (2.3) the electromagnetic Lagrangian
and using (2.12) the Hamiltonian becomes
$$H_{\rm RN}=p_a(bp_b-ap_a)-4b^2 + P_A^2 = H + P_A^2,\eqno(2.13)$$
where $P_A$ is the conjugate momentum to $A$. Since (2.13) is separable
we can solve the equation of motion for the electromagnetic field and
have $P_A=Q$ where $Q$ is the charge of the \bh. Eq. (2.11) becomes
$$H_{RN}= H + Q^2=0.\eqno(2.14)$$
The RN case is equivalent to the \Sc\ case with the constraint (2.14) in
place of (2.11).
\beginsection 3. Algebra and Gauge Transformations.
The gauge transformations of the system are generated by $H$ ($i=a,b$):
$$\eqalignno{&\delta q_i =\alpha (r) {{\d H} \over {\d p_i}} = \alpha(r) \bigl[
q_i,H \bigr]_P,&\hbox{(3.1a)}\cr\cr
&\delta p_i =-\alpha (r) {{\d H} \over {\d q_i}} = \alpha(r) \bigl[
p_i,H \bigr]_P,&\hbox{(3.1b)}\cr\cr
&\delta l = {{d \alpha} \over {dr}}.&\hbox{(3.1c)}\cr\cr}$$
The action (2.8) is invariant under (3.1) apart from a boundary term
that does not change the classical equations of motion:
$${\cal S}=\int_{r_1}^{r_2} dr {{d} \over {dr}} \biggl[ \alpha \bigl(
p_i {{\d H} \over {\d p_i}} + q_i{{\d H} \over {\d q_i}} -
2H\bigr) \biggr].\eqno(3.2)$$
With $\alpha \rightarrow l(r) dr$ eqs. (3.1a,b) are the equations of
motion.

The system described by the action (2.8) has remarkable algebraic
properties. Consider the following canonical quantities:
$$\eqalignno{&J = 8b - p_a p_b,&\hbox{(3.3a)}\cr\cr
&I = b/p_a.&\hbox{(3.3b)}\cr}$$
$J$ and $I$ are canonically conjugate gauge invariant quantities
(and also obviously integrals of the motion):
$$\bigl[J,I\bigr]_P = 1,~~~~\bigl[J,H\bigr]_P = 0,~~~~
\bigl[I,H\bigr]_P = 0.\eqno(3.4)$$
It is also interesting to consider the canonical quantity
$$N=bp_b - 2ap_a.\eqno(3.5)$$
We have
$$N = IJ+ 2H/p_a,\eqno(3.6)$$
and the relations
$$\bigl[N,H\bigr]_P = -2H,~~~\bigl[N,I\bigr]_P = -I,
{}~~~\bigl[N,J\bigr]_P = J.\eqno(3.7)$$
$N$ is not gauge invariant, however, in the case $Q^2=0$, i.e. for a
\Sc\ metric, it is constant on the constraint $H=0$. We shall see that
$N$ plays an interesting role in the frame of the WDW equation.

The gauge transformations (3.1) from a gauge $l_1$ to $l_2$ can be
integrated explicitly. We have
$$\eqalignno{&a=-H \alpha^2 + IJ \alpha + 4I^2,&\hbox{(3.8a)}\cr\cr
&b = -{{I} \over {\alpha}},&\hbox{(3.8b)}\cr\cr
&p_a = - {{1} \over {\alpha}},&\hbox{(3.8c)}\cr\cr
&p_b = J \alpha + 8 I,&\hbox{(3.8d)}\cr\cr
&\alpha = \int dr \bigl( l_2 - l_1\bigr).&\hbox{(3.8e)}\cr}$$
{}From eqs. (3.8) the gauge independent relation follows
$$a = {{I^2} \over {b^2}} \bigl( 4b^2 - Jb -H \bigr).\eqno(3.9)$$
On the constraint $H=0$ (\Sc\ metric)
$$a = 4I^2 \left( 1 - {{J} \over {4b}}\right). \eqno(3.10)$$
Therefore from (2.5)
$$ J = 8 M,\eqno(3.11)$$
where $M$ is the \Sc\ mass. On the constraint $H=-Q^2$ the two roots of
$a=0$ in (3.9) correspond to the two horizons of the RN metric.

It follows that in the case of the \Sc\ metric $I$ is the momentum
conjugate to the \Sc\ mass.  This suggests to perform a canonical
transformation to new pair of canonical variables, $(J,I;q_a,p_a)$ where
$q_a = -H/p_a^2$. This motivates our choice of the eigenfunctions in the
discussion of the WDW equation.
\beginsection 4. Quantization.
The quantization of this apparently simple system exhibits ambiguities
that are characteristic of the canonical quantization of systems
described by general relativity [11].

A main problem in general is that, in order to set up canonical
quantization rules, we must know a priori the causal structure of the
model representing a physical system. To be more specific, we must know
which coordinate plays the role of time and consequently write down equal
time canonical commutation relations.

This is usually an ambiguous procedure. In the classical treatment, the
identification -- if any -- of the time variable results from the
solution of the classical equations of motion and it is not determined a
priori. Of course, in some cases, as for instance the Friedmann --
Robertson -- Walker (FRW) model, one assumes the signature of the metric
(see e.g. [12]). This is because the outcome of the equations of motion
is anticipated, and a limitation in the signature of the metric is
consequently assumed. However, strictly speaking, these limitations are
not always known at the start. This becomes evident whenever the
classical equations allow for a change in the signature of the metric
(see e.g. [7]) or when, as in the present case, the presence of a horizon
induces a double change of signature in the metric.

In our present case we know from classical solutions that the signature
of the metric (and the gauge fixing of the coordinate) implies for $r$ a
timelike range. It is then tempting to explore the implications of a
canonical quantization of this system imposing equal $r$ commutation
relations. This will be carried out in the present section.

We shall impose the constraint (2.11) as an operator condition on the
wave function. This is the WDW equation. It expresses a necessary
condition for the wave function, although it does not in general contain
all the information relevant to the quantum form of the theory. Indeed as
it is well known the time is not identified, the solution contains both
positive and negative frequencies, it is a hyperbolic differential
operator and thus it does not lead to a well defined boundary value
problem. It is also plagued by ambiguities since the metric in the
Hilbert space is not defined.

We believe that the correct procedure [8] requires identification of the
parameter $r$ (our internal time) through a gauge fixing condition that
defines $r$ in terms of the canonical variables and leads to a unitary
Hamiltonian in the reduced canonical space. When this is possible the
quantization of the system is non ambiguous and the solutions contain
also the information from the constraint. The problem of the gauge fixing
in the present case will be treated elsewhere; here we shall limit
ourselves to explore the properties of the solutions of the WDW equation.

The fundamental commutation relations are:
$$\eqalignno{&\bigl[a,p_a\bigr] = i,&\hbox{(4.1a)}\cr
&\bigl[b,p_b\bigr] = i.&\hbox{(4.1b)}\cr}$$
The operators $a$, $p_a$, $b$, $p_b$ have the \Schr\ representation,
there being the usual ambiguities about the measure to be used.

We introduce also the mass operator $J$ and its conjugate $I$ according
to eqs. (3.4). We have
$$\bigl[ I,J\bigr] = i. \eqno(4.2)$$
We remark in particular that $J$ commutes with $p_a$.

The expression of the WDW Hamiltonian operator is (we consider for
simplicity the case of the \Sc\ metric, $Q=0$)
$$H_{WDW} = -ap_a^2 - bJ + 4b^2 + i \lambda p_a.\eqno(4.3)$$
The term $\lambda$ depends on the ordering  and on the representation of
$p_a$ and $p_b$. The choice of the covariant Laplace -- Beltrami operator
[13] leads to $\lambda=1$ while the symmetric ordering of the operators
$a,p_a$ and $b,p_b$ leads to  $\lambda=1/2$. In what follows we shall
keep $\lambda$ undetermined.

First of all we determine the eigenfunctions of the commuting operators
$J$ and $p_a$. We choose the simplest representation,
$$p_{q} \rightarrow -i {{\d}\over{\d q}},\eqno(4.4)$$
$(q=a,b)$. Then the eigenvalue equation for $J$ is
$$\bigl( 8b + \d_a \d_b \bigr)~ \psi_{M} = 8M~ \psi_{M},\eqno(4.5)$$
and the eigenfunctions of $J$ and $p_a$ are given by
$$\psi_{pM}(a,b) = \sqrt{{8\over p}} {{1} \over{2\pi}}
\exp~ i\left(pa +p^{-1} \beta_M \right),\eqno(4.6)$$
where $p$ is the eigenvalue of $p_a$ and
$$\beta_M = 4b(b-2M).\eqno(4.7)$$
The set (4.6) is orthonormal in $-\infty<a<+\infty$, $-\infty<b<+\infty$
with unit measure. Let us remark that this approach can be easily adapted
to a different interval in $a$, $b$ and to different representations for
$p_a$, $p_b$;  the wave function will change correspondingly, however the
important properties to exploit remain the role of $J$ and of the
commutation relation $[J,p_a]=0$.

The form of $\beta_M$ is related to the existence of the horizon at
$b=2M$ for positive $M$. Expressing the solution of the WDW equation
$$ H_{WDW} \Psi~ =0 \eqno(4.8)$$
as superposition of $\psi_{pM}$, the general representation of the WDW
wave function for the \Sc\ \bh\ is given by
$$ \Psi(a,b)~ =
\int dp~ p^{\lambda - 3/2} \int dm~ C(m)~ \psi_{pm}(a,b).\eqno(4.9)$$
$C(m)$ is  arbitrary. It is interesting to remark that there is a priori
no limitation on the sign of the mass $m$.

Using a well known representation for the solutions of the Bessel
equation [14], the representation (4.9) can be cast in the form
$$ \Psi(a,b)~ =\int dm~ C(m)  \left( -{{\beta_{m}}
\over{a}}\right)^{(\lambda-1)/2}~ K_{1-\lambda}(2 \sqrt
{-a \beta_m}) \eqno(4.10)$$
and the solution with fixed mass $M$ is
$$\Psi_{M}~ = C_M \left( -{{\beta_{M}}
\over{a}}\right)^{(\lambda-1)/2} K_{1-\lambda}(2 \sqrt
{-a \beta_{M}}). \eqno(4.11)$$
It is natural to assume the form (4.11) of the solution in the regions
where $a\beta_M<0$, namely in the classically forbidden regions
$a<0,~b>2M$ and $a>0,~b<2M$, where (4.11) is damped exponentially for
large $b$. In the two classically allowed regions for the \bh, namely
$b>2M$, $a>0$ and $b<2M$, $a<0$, the behaviour is oscillatory and one
should write the appropriate oscillating solutions with outgoing or
incoming asymptotic conditions. Note that for large $b$ in these regions
the phase approaches the value of the action evaluated on the classical
solution for the asymptotically flat spacetime. We are not discussing the
joining of the wave functions between the different regions as this
depends on the choice of the ordering and also on the representation
assumed for the momenta.

Suitable superpositions of the kind (4.10) may give
wave functions that are regular also for $b\rightarrow 0$ [10,15].
We note also that the general solution for the \ks\ Euclidean \wh\
found in [10] corresponds to the solutions of the present WDW equation
obtained by diagonalizing the operator $N$ (see (3.5)) in place of $J$.
Indeed, using the same choice for ordering ($\lambda=1$) and measure in
superspace as in [10], the differential representation of $N$ is
$$N=-i(b\d_b-2a\d_a),\eqno(4.12)$$
and the solutions of the WDW equation that are eigenfunctions of $N$ with
eigenvalue $\nu$ are:
$$\Psi_{\nu}(a,b)={8\over\pi}\left({2\sinh{\pi \nu}\over
\nu}\right)^{1/2}(-a)^{i\nu /2}K_{i\nu}\left(4b\sqrt{-a}\right).
\eqno(4.13)$$
These solutions are real in the region $a<0$ and orthonormal in $0\le b
\le\infty$, $-\infty\le a\le 0$ with measure $b~ da~ db$:
$$(\Psi_{\nu},\Psi_{\nu'})\equiv\int^0_{-\infty}da\int^{\infty}_0 db~b~
\Psi^*_{\nu}~\Psi_{\nu'}~=\delta(\nu-\nu').\eqno(4.14)$$
Again the phase factor coin\-ci\-des asym\-pto\-ti\-cal\-ly with the
clas\-si\-cal phase factor as for (4.11). It is interesting to note that
when $\nu=0$ the solution (4.13), namely
$$\Psi_{\nu=0}~ = {{8\sqrt{2}}\over{\sqrt{\pi}}}
K_0\bigl(4b \sqrt{-a}\bigr), \eqno(4.15)$$
coincides with (4.11) for $M=0$ (and $\lambda=1$), as expected since
$J\Psi_{M=0}(a,b)=N\Psi_{M=0}(a,b)=0$ on the constraint shell $H=0$. This
wave function describes a vacuum \wh\ in the classically forbidden
region. This equivalence supports the conjecture [16] that the ultimate
remnant in the evaporation process of a \bh\ is a vacuum \wh.
\beginsection 5. Conclusions.
The classical Einstein equations for a static spherically symmetric
metric can be cast in Hamiltonian form. The starting point is the ADM
foliation performed along the coordinate $r$. This is of course a
constrained canonical formalism, the constraint being that the
Hamiltonian vanishes. The Hamiltonian generates gauge transformations of
the canonical variables that correspond to the reparametrization of the
coordinate $r$ in the customary formalism of General Relativity.

By a suitable, self -- suggesting choice of the Lagrangian multiplier
(analogously to what done in [17,8] for the FRW universe) the Hamiltonian
assumes a beautiful polynomial form. The infinitesimal gauge
transformations can be integrated, thanks essentially to Einstein and
\Sc. This is an interesting integrable non linear system. Integrability
is due to its simple algebraic structure. Indeed, one identifies a pair
of conjugate gauge invariant quantities: one of them is the \Sc\ mass.

Then, the temptation to explore the quantization of this system is big
and we have carried on the investigation of the WDW equation. In doing
this, one is comforted by the fact that inside the horizon of a \bh\ $r$
is a timelike variable.

Note that if we do not fix the coordinate gauge by expressing $r$ in
terms of the canonical coordinates, this statement is vague: for instance
the trivially different fixings $b=r$ (area gauge) and $b=e^{r}$ lead to
obviously different values for the horizon in terms of $r$. However, this
does not matter much: there is a region where $r$ is timelike.

Thus we have studied the WDW equation and give the general representation
of the solution in terms of superpositions of eigenfunctions of the mass
operator. It is interesting to observe that there is no reason why the
sum should be limited to positive eigenvalues of the mass only.

There is nothing in the form of the WDW equation that reminds us of the
region in $a$, $b$ where it is valid, as the WDW equation does not
contain $r$. So we may determine the solution in the four regions
$a{>\atop <}0$, $b{>\atop <}2M$ (for positive mass). We have not
discussed the joining of the solutions between these regions, as the
result may be affected by the ambiguities in the ordering of the
operators and in the choice of the measure.

The solution in the classically forbidden regions can also be cast in a
form identical to the solution representing a Euclidean \wh\ in the
Kantowski -- Sachs spacetime [10]. These solutions are eigenfunctions of
a different operator $N$ that commutes weakly with the Hamiltonian. In
particular the state with eigenvalue 0 of $N$ is also eigenstate of the
mass with eigenvalue 0. This equivalence may support the conjecture [16]
that the ultimate remnant in the process of evaporation of a black hole
is a vacuum \wh.

The WDW equation is plagued by the so well known problems. A more natural
way to investigate the quantum properties of the system seems to be the
introduction of a gauge fixing of the parameter in the canonical
treatment [8] that connects $r$ to the canonical variables and leads to a
unitary Hamiltonian in the reduced canonical space. We defer to a next
paper the investigation of this method as well as of the connection
between the WDW equation and the gauge fixed quantization for integrable
systems.
\beginack
It is a pleasure to thank Orfeu Bertolami, Fernando de Felice and Luis
J. Garay for interesting discussions on the subject of this paper and
related topics. One of the authors (A.T.F.) acknowledges a partial
support from the Russian Science Foundation (grant 93 - 02 - 3827) and
from the International Science Foundation (grant RF 000).
\smallskip
\beginref
\def\Kk{H.A. Kastrup and T. Thiemann, \NP {\bf B425}, 665 (1994);
K.V. Ku\-cha\v r, \PRD {\bf 50}, 3961 (1994).}

\def\Wil{P. Kraus and F. Wilczek, {\tscors Some Applications
of a Simple Stationary Line Element for the \Sc\ Geometry}, Report
No: PUPT 1474, IASSNS 94/46, gr-qc/9406042.}

\def\Bra{S.P. Braham, {\tscors Hypertime Formalism for Spherically
Symmetric \BHs\ and \Whs}, Report No: QMW-Maths-1994-SPB-1,
gr-qc/9406045.}

\def \Wei{See for instance: S. Weinberg, {\tscors Gravitation and
Cosmology: Principles and Applications of the General Theory of
Relativity}, John Wiley \& Sons, New York, 1972.}

\def\Kkk{For a review of quantum gravity problems, see for instance:
K. V. \K, {\it Time and Interpretations of Quantum Gravity} in {\it
Proc. 4th Canadian Conference on General Relativity and Relativistic
A\-stro\-phy\-sics}, World Scientific, Singapore, 1993.}

\def\Whe{J.A. Wheeler, in {\tscors Battelle Rencontres: 1967 Lectures
in Mathematics and Physics}, eds. by C. DeWitt and J.A. Wheeler,
Benjamin, New York, 1968.}

\def\Dew{B.S. DeWitt, \PRD {\bf 160}, 1113 (1967).}

\def\Mar{J. Martin, \PRD {\bf49}, 5086 (1994).}

\def\Cdf{M. Cavagli\`a, V. de Alfaro and A.T. Filippov, {\tscors
A Schr\"odinger Equation for Mini Universes}, Report No: DFTT 6/94,
SISSA 25/94/A, gr-qc/9402031, \IJMPA in press.}

\def\Haw{S.W. Hawking and D.N. Page, \PRD {\bf 42}, 2655 (1990).}
\def\Cav{M. Cavagli\`a, \MPLA {\bf 9}, 1897 (1994).}

\def\Haww{S.W. Hawking, \PRL {\bf 69}, 406 (1992).}

\def\Bb{Bateman Manuscript Project, {\tscors
Higher Trascendental Transforms, Vol. II}, p. 82,
Mc. Graw--Hill Book Company, New York, 1953.}

\def\Ell{G. Ellis, A. Sumeruk, D. Coule, and C. Hellaby, \CQG {\bf
9}, 1535 (1992).}

\def\Cdd{M. Cavagli\`a, V. de Alfaro, F. de Felice, \PRD {\bf 49}, 6493
(1994).}

\def\Cd{M. Cavagli\`a and V. de Alfaro, \MPLA {\bf 9}, 569 (1994).}

\def\Gar{See for instance: L.J. Garay, \PRD {\bf 48}, 1710 (1993); G. A.
Mena Marugan, \CQG {\bf 11}, 2205 (1994) and \PRD {\bf 50}, 3923 (1994).}

\item{[1]} \Kk
\item{[2]} \Wil
\item{[3]} \Bra
\item{[4]} \Wei
\item{[5]} \Whe
\item{[6]} \Dew
\item{[7]} \Mar
\item{[8]} \Cdf
\item{[9]} \Cdd
\item{[10]} \Cav
\item{[11]} \Kkk
\item{[12]} \Ell
\item{[13]} \Haw
\item{[14]} \Bb
\item{[15]} \Gar
\item{[16]} \Haww
\item{[17]} \Cd

\endref
\bye